\documentclass[%
 reprint,
 superscriptaddress,
 nofootinbib,
 amsmath,amssymb,
prx,
showkeys,
longbibliography
]{revtex4-2}
\usepackage{graphicx}
\usepackage{amsmath}
\usepackage{amssymb}
\usepackage{ulem}
\usepackage{xcolor}

\begin{document}
\title{Analytical solution for the polydisperse random close packing problem in 2D}





\author{Alessio Zaccone}
\affiliation{Department of Physics ``A. Pontremoli'', University of Milan, via Celoria 16, 20133 Milan, Italy.}

\begin{abstract}
An analytical theory for the random close packing density, $\phi_\textrm{RCP}$, of polydisperse hard disks is provided using an equilibrium model of crowding [A. Zaccone, Phys. Rev. Lett. 128, 028002 (2022)] which has been justified on the basis of extensive numerical analysis of the maximally random jammed (MRJ) line in the phase diagram of hard spheres [Anzivino et al., J. Chem. Phys. 158, 044901 (2023)]. 
The solution relies on the equations of state for the hard disk fluid and provides predictions for $\phi_\textrm{RCP}$ as a function of the ratio, $s$, of the standard deviation of the distribution of disk diameters to its mean. 
For a power-law size distribution with $s=0.246$, the theory yields $\phi_\textrm{RCP} =0.892$, which compares well with the most recent numerical estimate $\phi_\textrm{RCP} =0.905$ based on the Monte-Carlo swap algorithms [Ghimenti, Berthier, van Wijland, Phys. Rev. Lett.  133, 028202 (2024)].

\end{abstract}

\maketitle

The hard-sphere (HS) system is a standard paradigm in statistical physics, and has proved invaluable in understanding the microscopic structure of condensed matter, in particular liquids and crystals.
The absence of interactions apart from a hard-core repulsion makes it amenable to analytical treatments to accurately describe its equation of state, and allows precise numerical calculations of its phase behavior.
The latter is governed only by the filling fraction $\phi$ and not by a temperature, unlike other standard models of statistical mechanics.
An equilibrium, compression of the HS fluid leads to an entropy-driven first-order freezing phase transition~\cite{alder1957phase, wood1957preliminary, hoover1968melting}, a paradigmatic phenomenon that catches the essential features of freezing in most simple liquids~\cite{Hansen2006}.
Eventually, compression leads to a close-packed configuration that optimally fills space at close-packing (CP), $\phi_{\textrm{CP}}$.
If compression is applied rapidly to the system, equilibrium freezing is avoided, and the largest achievable density shifts to a value lower than $\phi_{\textrm{CP}}$ as particles randomly ``jam'' in a disordered configuration, accompanied by the divergence of pressure \cite{van1995real,sanz2011crystallization,zaccarelli2009crystallization}.
The maximum packing fraction achieved in this disordered state is known as the random close packing (RCP) volume fraction, $\phi_{\textrm{RCP}}$ \cite{bernal1960packing, zaccone2022explicit, anzivino2023estimating, zaccone2023}.
All these facts hold for both monodisperse and polydisperse spheres in $d=3$.

In $d=2$, the physics of hard disks presents some fundamental differences. While the Mermin-Wagner-Hohenberg argument appears to rule out long-range order in 2D, the first computer simulations showed that the hard disk fluid undergoes crystallization \cite{alder1957phase}. 
The work of Berezinskii, Kosterlitz and Thouless (BKT) \cite{Berezinski_1,Berezinski_2,Kosterlitz} was motivated by this apparent contradiction and conclusively showed the existence of a crystalline phase also in 2D, with quasi-long range order, \textit{i.e.} power-law decay of spatial correlations. 
The freezing of equilibrium monodisperse hard disks occurs via a two-step mechanism: first a liquid-hexatic transition around $\phi\approx 0.72$, closely followed by a transition from the hexatic phase to a triangular crystal.
The original BKT theory predicts that both transitions are continuous, in agreement with the first experimental verification of the BKT scenario using superparamagnetic colloids interacting via a long-range dipolar potential \cite{Keim1,Keim}. However, subsequent evidence from numerical simulations suggests that the liquid-hexatic transition is first-order \cite{Krauth,Dijkstra,Tsiok}, while more complex scenarios have been reported for the 2D melting of multi-component mixtures \cite{Pica}.

In $d=2$, a broad range of $\phi_{\textrm{RCP}} \approx 0.81-0.89$ has been reported using different algorithms and theories \cite{zaccone2022explicit,Stillinger,Sutherland,Brouwers,Brouwers2024,Sugiyama,Berryman,Makse}, the upper bound of which is quite close to the CP value, $\phi_\textrm{CP} =\pi/\sqrt{12} \approx 0.9069$ (for comparison, in 3D, $\phi_{\textrm{RCP}} \approx 0.64-0.65$ and $\phi_\textrm{CP} =\pi/\sqrt{18} \approx 0.7404$).
This proximity between $\phi_\textrm{CP}$ and $\phi_{\textrm{RCP}}$ makes it harder to entirely avoid crystallization, to the point that some methods to generate disordered packings miss it altogether in 2D~\cite{Wilken2023}.

Importantly, as pointed out in \cite{truskett2000towards}, while
$\phi_{\textrm{CP}}$ is unambiguously defined as the largest achievable filling fraction,
$\phi_{\textrm{RCP}}$ is not.
In particular, slight changes to the compression procedure may change the observed range of values~\cite{Ozawa2017}.
Yet, one may introduce well-defined special packings, that are most disordered for a given choice of order parameter, \textit{e.g.} the bond-orientational order parameter~\cite{truskett2000towards}, which leads to the well-defined concept of maximally random jammed (MRJ) states~\cite{torquato2010jammed}.

Alternatively, in fast compressions, RCP is the lowest volume fraction at which the static (low-frequency) shear modulus rises from zero.
For the jamming of soft repulsive particles, the non-affine analytical theory of the shear modulus value shows that this is tantamount to requiring that RCP is isostatic, meaning that the average number of contacts $z$ of a particle at RCP verifies $z = 2d$~\cite{zaccone2011approximate}, in agreement with numerical observations with various dynamics~\cite{OHern,Ozawa2017,wilken2021random}.
Defining RCP as the densest isostatic packing, an analytical estimate for $\phi_{\textrm{RCP}}$ can be obtained for monodisperse hard spheres in both 2D and 3D, using an approximate description of crowding along the line linking RCP to CP~\cite{zaccone2022explicit,anzivino2023estimating}.
The underlying assumption is that crowding along the family of least coordinated packings can be approximated by a rescaled equilibrium HS theory, as justified through numerical simulations in Ref.~\cite{anzivino2023estimating}.
This approach was extended to predict the $\phi_{\textrm{RCP}}$ for bidisperse and polydisperse hard spheres, as a function of the mean, standard deviation, and skewness of the particle size distribution, using the available equilibrium models for bidisperse and polydisperse hard spheres as the input \cite{anzivino2023estimating}.
This estimate for polydisperse random close packing has found widespread application \cite{Lombard,Hilgenfeldt,Jaeger,Malamud} because it allows one to estimate the density of a granular medium or powder based on the first three moments of the particle size distribution (PSD).
For example, it has been used to estimate the density of the lunar regolith based on the standard deviation of the PSD measured on the samples brought back by the Apollo mission \cite{Blum,Yu_2024}.
Furthermore, understanding the maximum packing density of polydisperse particles is essential for tailoring the properties of a number of materials, from reinforced polymer nano-composites \cite{Bonn} to the microstructure and properties of polycrystalline and amorphous materials \cite{Krief,Zaccone_2020}.

Following the same approach, here we present the analytical solution for the random close packing of polydisperse particle monolayers or random close packing of disks in 2D~\cite{meer2024estimating}. To date, no analytical solution to this problem has been reported in the literature. The 2D polydisperse close packing also presents important connections with various applications, from particle monolayers at interfaces \cite{Isa,Maestro,Lazar_2024,Keten,Seto}, to the molecular packing of 2D polymer films \cite{Laso,Strano}, to self-assembled microelectronic materials \cite{Milani,An2024}, and 2D cellular and organelles assemblies in biology \cite{schramma2025,Julicher,Gov,Heisenberg,Merkel}.

For a 2D system of polydisperse particles, the mean number of contacts, $z_{ij}$, between particles of species (size) $i$ and those of species (size) $j$ is linked to the partial radial distribution function (rdf), $g_{ij}(r)$, restricted to $ij$ pairs, via the following general relation 
\begin{equation}
z_{ij} = 2 \pi \rho \int_{0}^{\sigma_{ij}^+} \mathrm{d}r r g_{ij}(r),
\end{equation}
where only the metric factor in the integration differs from the 3D case~\cite{anzivino2023estimating}.
Here, $\sigma_{ij}^+ \equiv \sigma_{ij} + \epsilon$, with $\epsilon \rightarrow 0$.

By taking an average over all $ij$ contacts, this becomes:
\begin{equation}
z = 2 \pi \rho \int_{0}^{\sigma^+} \mathrm{d}r r g(r), \label{coord}
\end{equation}
where $z$ is the mean contact number across the entire system, $\sigma$ the mean diameter, $\rho=N/A$ the number of disks per unit area, and $g(r)$ the total rdf averaged over all $ij$ contacts.

In the polydisperse case \cite{anzivino2023estimating}, like in the monodisperse case \cite{zaccone2022explicit}, the rdf can be described as a partially-continuous distribution, with a discrete-like contact part $g_c(r)$ and a continuous ``beyond contact'' (BC) part $g_{BC}(r)$ \cite{Torquato_JCP}:
\begin{equation}
    g(r) = g_c(r) + g_{BC}(r)
\end{equation}
where $g_c(r)=g_0 g(\sigma;\phi)\delta(r-\sigma)$ \cite{zaccone2022explicit,Torquato_JCP}, with $g(\sigma;\phi)$ the contact value of the averaged rdf,  and $g_0$ is a dimensional constant (with dimension $1/length$) to be determined via a suitable boundary condition.
Plugging this form of $g(r)$ into Eq.~\ref{coord}, we get
\begin{equation}
    z=8 \, \phi \, \frac{g_0}{\sigma}g(\sigma;\phi) \label{pivo}
\end{equation}
where the filling fraction $\phi$ is defined as $\phi = {\pi N \sigma^2}/({4 S})$, where $S$ is the total surface area.

In an equilibrium fluid the compressibility $Z(\phi)$, as given by an equation of state (EOS), is related to the contact value , $g_{eq}(\sigma;\phi)$ of the rdf by means of the virial theorem \cite{Ree}
\begin{equation}
Z(\phi) \equiv P/\rho k_B T = 1+ B_2 \rho g_{eq}(\sigma;\phi)
\end{equation}
where $P$ is the pressure and $B_2 = \pi \sigma^2/2$ the second virial coefficient. This relation can be simplified as
\begin{equation}
    Z(\phi) = 1 + 2 \phi g_{eq}(\sigma;\phi).
\end{equation}
By analogy with this equilibrium result, and as justified via numerical simulations in Ref.~\cite{anzivino2023estimating} for the 3D case, we introduce the approximation 
\begin{equation}
    g(\sigma;\phi) \propto \frac{Z(\phi)-1}{2\phi}. \label{cont}
\end{equation}

By combining Eq. \eqref{pivo} with Eq. \eqref{cont}, we arrive at the fundamental relation which will be pivotal to estimate $\phi_{\textrm{RCP}}$:
\begin{equation}
     z=4 \,C_0 [Z(\phi)-1], \label{fund}
\end{equation}
where we introduced a dimensionless proportionality constant $C_0$.
As explained in previous work \cite{zaccone2022explicit,anzivino2023estimating,Likos}, $C_0$ can be evaluated by pinning the $z(\phi)$ curve to the CP value for a monodisperse packing, which for 2D disks corresponds to triangular close-packing $\phi_{CP}=\pi/\sqrt{12}\approx 0.9069$ and $z_{ref} \equiv z(\phi_{CP}) = 6$.

The predicted values depend on the choice of EOS $Z(\phi)$ used to describe mono- and polydisperse disks.
Many expressions were proposed over the years for the monodisperse EOS, with varying degrees of analytical and numerical groundings~\cite{mulero2008theory}.
In this paper, we use three common choices due to their relative simplicity, namely the scaled-particle theory EOS ($Z_{SPT}$) \cite{Helfand}, its modified version proposed by Henderson ($Z_H$) \cite{Henderson}, and the 2D version of the Carnahan-Starling expression ($Z_{CS}$) \cite{CS}.
These choices of EOS yield similar values for $C_0$ when substituting the relevant $Z(\phi_{ref})$ in Eq. \eqref{fund},
\begin{equation}
C_0 = \frac{z_{ref}}{4(Z(\phi_{ref})-1)},\label{const}
\end{equation}
as shown in table~\ref{tab:const_table}.
\begin{table}[]
    \centering
    \begin{tabular}{|c|c|c|}
        \hline
         EOS & $C_0$ & $\phi_{\textrm{RCP}}$  \\
         \hline
         $Z_{SPT}(\phi) = 1/(1-\phi)^{2}$ & 0.0131152 & 0.886222\\
         \hline
         $Z_H (\phi) = (1 + \phi^2 / 8) / (1-\phi)^{2}$ & 0.0118828 & 0.886436 \\
         \hline 
         $Z_{CS}(\phi) \approx (1 - 0.43599 \phi) / (1-\phi)^{2}$ & 0.0218171 & 0.885509 \\
         \hline
    \end{tabular}
    \caption{\textbf{Predictions for monodisperse EOS.}
    We report the expression of the EOS, and the corresponding numerical values of $C_0$ and the estimate of the monodisperse $\phi_{\textrm{RCP}}$.}
    \label{tab:const_table}
\end{table}

To extend the EOS to the polydisperse case, we follow the approach of Ref.~\cite{Ogarko}, which is valid in the high-density regime of equilibrium disks.
This methodology is based on a fundamental-measure-theory (FMT) approach that establishes a connection between  the EOS of a monodisperse HS fluid and that of a polydisperse HS mixture \cite{Santos}.
In essence, this approach posits that the excess free energy of the polydisperse mixture at a given packing fraction $\phi$ can be mapped onto that of a single-component fluid at a different effective packing fraction, $\phi_{\textrm{eff}}$.
As derived by Santos and co-workers \cite{Santos_persp,Santos,Ogarko}, the relation between $\phi_{\textrm{eff}}$ and $\phi$ reads as follows:
\begin{equation}
    \phi_{\textrm{eff}} = \frac{\phi}{\phi + \lambda {(1-\phi)}}.
\end{equation}
Here, $\lambda$ is a number that depends on the second and third moments of the PSD:
\begin{equation}
    \lambda = \frac{m_3}{m_2^2}
\end{equation}
where $m_q \equiv M_q/M_1^q$ is the $q$th dimensionless moment of the PSD, and $M_q = \int_0^{\infty} \sigma^q p(\sigma) d\sigma$ is the $q$th moment, with $p(\sigma)$ the normalized density function of the PSD. 

The compressibility of the polydisperse mixture as derived by Santos and co-workers \cite{Ogarko} reads as:
\begin{equation}
   Z(\phi)= \frac{1}{1-\phi} + \frac{\alpha}{\phi} \left[ \phi_{\textrm{eff}} Z_s(\phi_{\textrm{eff}}) -\frac{\phi_{\textrm{eff}}}{1-\phi_{\textrm{eff}}} \right] \label{compr_fin}
\end{equation}
where $Z_s(\phi_{\textrm{eff}})$ is the compressibility of the monodisperse system, and $\alpha = \lambda / m_2$.

The random close packing fraction $\phi_{\textrm{RCP}}$ of disks can thus be obtained by requiring that RCP occurs at the rigidity onset, \textit{i.e.} at the isostatic point $z_c=4$ in Eq. \eqref{pivo} \cite{zaccone2011approximate}, leading to:
\begin{equation}
    C_0 [Z(\phi_{\textrm{RCP}})-1]=1,
    \label{final}
\end{equation}
with $Z(\phi)$ given by Eq. \eqref{compr_fin} and $C_0$ given by Eq. \eqref{const}.

\begin{figure}[t]
    \centering
    \includegraphics[width= 0.9 \linewidth]{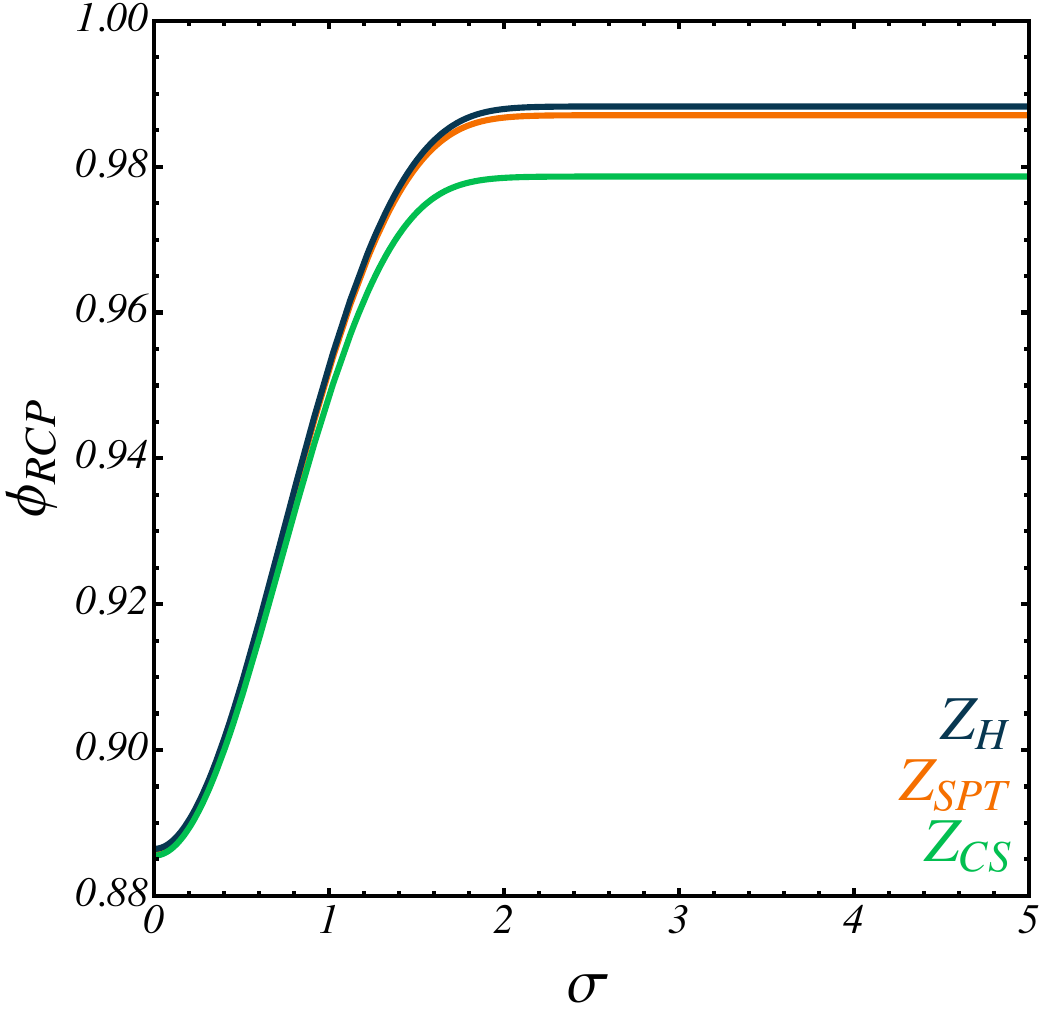}
    \caption{\textbf{Estimate for a lognormal size distribution.}
    $\phi_{\textrm{RCP}}$ of hard disks computed according to Eq. \eqref{final} using the $Z_H$ (blue), $Z_{SPT}$ (orange), and $Z_{CS}$ (green) EOS, and a log-normal particle size distribution.}
    \label{fig:lognormal}
\end{figure}

Both $\lambda$ and $\alpha$ depend on the first three moments $m_{1,2,3}$ of the PSD only through the rescaled standard deviation $s = \sqrt{m_2 - 1}$ and the reduced skewness $t = \sqrt[3]{m_3 - 3 m_2 + 2}$, so that Eq. \eqref{final} can be solved to obtain $\phi_{\textrm{RCP}}$ as a function of the parameters of an arbitrary size distribution.
In practice, the moments are usually not independent and are functions of simpler parameters that tune the distribution of sizes.
For instance, for a log-normal PSD,
\begin{align}
    p(x) = \frac{1}{x \sqrt{2\pi \sigma^2}}e^{-(\ln x- \mu)^2 / (2 \sigma^2)},
\end{align}
where $\mu$ and $\sigma$ are the mean and standard deviation of the underlying normal distribution. By employing the Santos EOS for $Z(\phi)$, the RCP density becomes solely dependent on a single free parameter, $\sigma$.
We report our predictions for $\phi_{\textrm{RCP}}$ against $\sigma$ in Fig.~\ref{fig:lognormal}.

We show that our estimate predicts a growth of $\phi_{\textrm{RCP}}$ with the width of the particle size distribution, until a saturation at $1/(1+C_0) \approx 0.97-0.98$~\cite{anzivino2023estimating}, a value that is, reassuringly, always below $1$.
The main difference between choices of EOS is the value of the saturating density at large $\sigma$, but all expressions yield similar outputs at moderate polydispersity.

The predictions of our approach can be compared with the most recent numerical estimates of the RCP packing fraction for random close packing of polydisperse hard disks obtained by irreversible Swap-Monte Carlo algorithms \cite{Ghimenti}.
In that work, the authors use a size distribution generated by gradient descent in the space of particle positions and (constrained) radii, starting from a log-normal distribution~\cite{Hagh2022,Bolton-Lum2024}, which remains nearly unchanged through the process.
The goal of the technique is to achieve the highest packing densities possible while maximizing the disorder.
In particular, irreversible Monte Carlo algorithms have been shown to successfully accelerate the sampling by breaking detailed balance~\cite{Bernard2009}, thus outperforming previous algorithms in preparing very dense jammed packings of disks~\cite{Ghimenti}.
In Ref. \cite{Ghimenti}, the maximum achievable $\phi_{\textrm{RCP}}$ for a polydisperse system with fixed reduced standard deviation $s=0.25$, corresponding to a log-normal $\sigma \approx 0.2462$, was estimated to be $\phi_{swapMC}\approx 0.905$.
This value can be compared with the analytically predicted value $\phi_{\textrm{RCP}}=0.8924$ using the above theory in conjunction with the EOS for polydisperse disks by Santos et al. \cite{Ogarko} (and the Henderson EOS \cite{Henderson} for $Z_s(\phi_{\textrm{eff}})$).
Note that the packings obtained in Ref.~\cite{Ghimenti} are not necessarily strictly isostatic (viz., $z = 4$); indeed, one might expect a larger number of contacts. As a result, our prediction is likely an underestimate of their final packing fraction, consistent with the reported value.

In conclusion, we have presented an analytical solution for the random close packing of polydisperse hard disks.
The solution is based on identifying RCP with the onset of rigidity in a crowding HS fluid \cite{zaccone2022explicit,anzivino2023estimating}, whereby the contact value of the rdf is assumed to be proportional to the equilibrium value predicted, for hard disks, by the approach of Santos and co-workers \cite{Ogarko}.
The theory predicts a monotonically increasing RCP volume fraction as a function of the reduced standard deviation, $s$, of the particle size distribution, which eventually saturates to a plateau $\phi_{\textrm{RCP}}\approx 0.97-0.98$ as $s \to \infty$.
At finite polydispersity, $s=0.25$, the analytical theory provides a value of RCP density that is in good agreement with those predicted by the most recent irreversible Monte Carlo algorithms achieving extremely dense jammed packings via collective swaps \cite{Ghimenti}.

\section*{Data availability}
No new data were created or analysed in this study.

\subsection*{Acknowledgments}
I am indebted to Dr. Mathias Casiulis for having thoroughly checked all the steps in the derivation and for assistance with Fig. 1.
Many useful discussions with Dr. Carmine Anzivino and with Prof. Stefano Martiniani are gratefully acknowledged.
A.Z. gratefully acknowledges funding from the European Union through Horizon Europe ERC Grant number: 101043968 ``Multimech'', from US Army Research Office through contract nr. W911NF-22-2-0256, and from the Nieders{\"a}chsische Akademie der Wissenschaften zu G{\"o}ttingen in the frame of the Gauss Professorship program. 

\bibliographystyle{apsrev4-1}

\bibliography{references}
\end{document}